\documentclass{sig-alternate}
\begin{document}
\conferenceinfo{EC'00,} {October 17-20, 2000, Minneapolis, Minnesota.}
\CopyrightYear{2000}
\crdata{1-58113-272-7/00/0010}

\newtheorem{theorem}{Theorem}
\newtheorem{lemma}{Lemma}
\newtheorem{proposition}{Proposition}
\newtheorem{definition}{Definition}
\newtheorem{example}{Example}
\newtheorem{corollary}{Corollary}

\title{Optimal Solutions for Multi-Unit Combinatorial Auctions:
Branch and Bound Heuristics}

\numberofauthors{2}
\author{
\alignauthor Rica Gonen\\
	\affaddr{School of Computer Science and Engineering}\\
	\affaddr{Hebrew University, Jerusalem 91904, Israel}\\
	\email{rgonen@cs.huji.ac.il}
\alignauthor Daniel Lehmann\\ 
	\affaddr{School of Computer Science and Engineering}\\
	\affaddr{Hebrew University, Jerusalem 91904, Israel}\\
	\email{lehmann@cs.huji.ac.il}
}

\maketitle
\begin{abstract}
Finding optimal solutions for multi-unit 
combinatorial auctions is a hard problem
and finding approximations to the optimal solution is also hard.
We investigate the use of Branch-and-Bound techniques:
they require both a way to bound from above the value of the
best allocation and a good criterion to decide which bids are to be
tried first.
Different methods for efficiently bounding from above the value of the
best allocation are considered.
Theoretical original results characterize the best approximation ratio
and the ordering criterion that provides it.
We suggest to use this criterion. 
\end{abstract}

\keywords{Combinatorial Auctions, Branch and Bound
}
\section{Multi-unit combinatorial auctions (MUCAs)}
\label{sec:MUCAs}
Auctions have been used from times immemorial, but the renewed 
modern interest in auctions stems from:
\begin{itemize}
\item their increased use for selling off government property after WWII
and later in extensive denationalizations, and
\item the theoretical breakthroughs started by~\cite{Vickrey:61}.
\end{itemize}
A very recent surge of interest in auctions stems from their surprising
success on the internet. 
Many foresee internet auctions in which:
\begin{itemize}
\item the number of buyers will be large
\item the number of items for sale will be large
\item the mechanism used to determine the allocation and the payment will
be complex both from the game-theoretic strategic and from
the computational points of view.
\end{itemize}
This work deals with computational aspects of winner determination in
multi-unit combinatorial auctions. 
Combinatorial auctions are mechanisms in which a number of items
are up for sale and bidders bid for subsets of those items.
In combinatorial auctions one typically assumes that all items for sale
are different.
Computational aspects of combinatorial auctions have been considered 
in~\cite{Rothkopf:9509,Sandholm99,FLBS:99,LCS:EC99,Nisan:Bidding,Tenn:sometractauc}.
Multi-unit combinatorial auctions are those combinatorial auctions 
in which certain of the items for sale are identical.
We assume $n$ different commodities and $k_{j}$ identical items of commodity
$j$ for \mbox{$j=1, \ldots n$}.
The problem we are trying to solve is that of determining the optimal
allocation of the items.
A set of bids is given: each bid requests a number of units (possibly zero) 
from each commodity and offers a price $p$ for the whole set.
A subset of the set of all bids is conflict-free if, for each commodity,
the sum of the units requested does not surpass the number of units for sale.
The problem is to find a conflict-free subset that maximizes
the sum of the prices proposed. 
Multi-unit combinatorial auctions have been considered in~\cite{Camus2:unpub}.
\section{Branch-and-Bound search}
\label{sec:BB}
The problem of finding the optimal solution of a multi-unit combinatorial
auction is a hard problem, i.e., requires extensive resources, as
will be shown formally in Section~\ref{sec:theory}.
It is therefore essential that one designs carefully the algorithm and
uses the available resources smartly.
The method we propose to use is based on the 
branch-and-bound search technique developed by operations 
researchers~\cite{LawlerWood:66}.
The use of some branch-and-bound techniques for finding optimal solutions
for combinatorial auctions has been considered in~\cite{FLBS:99} 
and more thoroughly in~\cite{Nisan:Bidding}. 
The case of multi-unit combinatorial auctions is considered 
in~\cite{Camus2:unpub} where a wealth of heuristics are proposed.
This paper concentrates on the fundamentals and characterizes the best
ordering heuristics.

The general description of the branch-and-bound technique below is well-known.
The resource in shortest supply is space (i.e. memory) and therefore 
we propose to use a depth-first search, with backtracking to cover the
whole search space.
At any time we essentially keep in memory only a partial solution, 
i.e., a set of non-conflicting bids. 
We look for a bid that can be added to the partial solution without
creating a conflict and if there is none, know we have attained a 
possible solution and backtrack.
The additional information of which we need to keep track consists only of the
best solution found so far and backtracking information. On the whole
the space requirements are linear in the size of the problem.
Note that, at any time, we have allocated part of the items and
are left with a smaller problem of the same type: finding the optimal
solution of another, smaller, MUCA.

An exhaustive search such as sketched above will require exponential time
for all possible inputs. One may hope to find an algorithm that will
run faster on the {\em easy} inputs. 
Even though the theoretical considerations
of Section~\ref{sec:theory} seem to preclude that there be a majority
of {\em easy} cases, one may hope that many of the problems that
will have to be solved in practice will turn out to be easy.
We are hoping for an algorithm that will run fast on those easy problems.
The powerful {\em branch-and-bound} idea is that, in many situations,
we may be able to conclude on the spot that adding a specific bid to our
partial solution is hopeless and that we can immediately backtrack,
saving us the exploration of a whole sub-tree, i.e., enabling us to {\em prune}
a whole sub-tree. 
All we need for that is a
good upper-bound for the value of the optimal solution of the smaller
MUCA concerning the items still unallocated.
If we have such an upper bound and if the sum of this upper bound and
the values of the bids of our partial solution is not larger than the
value of the best solution found so far, we may backtrack immediately:
no extension of our partial solution may have a value larger than
that of the solution we already know.
In Section~\ref{sec:bounding}, we shall propose a number of ways
to bound from above the value that can be obtained from auctioning
the still unallocated goods. There is obviously no need to commit
to a single method: one may use many such methods, obtain a number
of different upper bounds and use the smallest of all.

The pruning described above is most effective if the best solution
found so far, that is used to decide whether to prune or not, 
is in fact a good solution. 
If we find the optimal (or a very good) solution early on,
even though we shall not know that it is the optimal solution 
(or how good it is), we shall be able to prune more sub-trees than
if we had a best solution so far that is not as good.
It is therefore essential that we find the best solutions early on.
We should, therefore, carefully choose the order in which we try
to enter the bids into partial solutions and try the most promising
bids first. 
To summarize, we need a good heuristic to pick the most promising
bid in a set of bids, and this is the third component of branch-and-bound
method.
We shall use such a heuristic to find the most promising bid in the
smaller MUCA concerning the items still unallocated iteratively to
find the most promising of those bids that does not conflict with
the partial solution at hand. We shall then use the upper-bound
described above to decide whether to prune the whole sub-tree or not.
Notice that we do not propose to rank the bids once and for all:
the most promising bid may depend on the partial solution at hand.
For example, the normalized criteria described below in 
Equations~\ref{eq:euclidnorm} and~\ref{eq:norm} imply such a dependency.
In~\cite{Camus2:unpub} the heuristic proposed to choose the next bid
takes into account the whole set of bids.
We restrict our attention to heuristics that consider each bid separately.
In Section~\ref{sec:ordering} we shall consider different criteria
for this task and in Section~\ref{sec:theory} we shall characterize 
the best one. It turns out that this optimal criterion does not
depend on the partial allocation, or, equivalently, on the set of
items still unallocated, and therefore one can simply order the bids
according to this criterion once and for all at the start of the
algorithm.

In this paper, we shall assume that any subset of bids such that,
for any commodity, the sum of the units requested does not exceed the
total number of units available is a legal solution.
We shall not consider explicitly the case in which certain bids
are exclusive, i.e., no legal solution can contain certain pairs of bids.
The generalization of our (and essentially any) 
branch and bound algorithm to the case
of possibly exclusive bids is quite straightforward, though:
when considering whether to add a specific bid to a partial allocation,
check not only whether there are enough units left to satisfy it
and whether the sub-tree can be pruned, check also whether the bid
considered is not excluded by some bid of the partial solution.
\section{Bounding from above}
\label{sec:bounding}
We noticed that, at any point, we have a partial allocation,
i.e. a set of non-conflicting bids, and are trying to extend it.
The best possible extension is, again, the optimal solution to
the MUCA of the remaining units to the remaining bids.
We shall therefore look for ways to bound from above the value
of any MUCA.
We also noticed that one may use a number of different methods,
obtain numerous upper-bounds, and take the smallest one.
We propose three different types of methods.
\subsection{Linear Programming}
\label{subsec:LP}
The first method we propose is Linear Programming.
Our problem is an integer-programming problem: find 
\mbox{$x_{i} \in$} \mbox{$\{ 0 , 1 \}$} for each bid $i$ that maximizes
\mbox{$\sum_{i} x_{i} \: p(i)$} while satisfying the linear constraints:
\mbox{$\sum_{i} x_{i} \: q(i,j) \leq k_{j}$}.
The variable $x_{i}$ indicates whether bid $i$ is in the optimal
solution (\mbox{$x_{i} = 1$}) or out of it (\mbox{$x_{i} = 0$}).
The relaxed linear programming problem in which we allow
\mbox{$x_{i} \in [ 0 , 1 ]$} will provide a value
\mbox{$\sum_{i} x_{i} \: p(i)$} that is an upper-bound for the value
of the optimal solution to the original MUCA.
Notice that the solution of the relaxation allows for fractional
allocation of bids: \mbox{$x_{i} = 1/2$} may be interpreted as
allocating half of the quantities requested by $i$ at half the
price proposed.
In the worst case, for theoretical reasons, it seems that the upper bound
provided by LP cannot be good, but there are reasons to think that
in practice, and especially for large $k_{j}$'s the bound could
be pretty good quite often. An extended discussion of the relation
between the solutions to the original and the relaxed problems may
be found in~\cite{Nisan:Bidding}.

Notice also that, if the optimal solution to the relaxed
problem is integral, then we know it is the optimal solution
to the original problem. The integrality of the solution to the LP
problem is the signal for backtracking.
\subsection{Projections}
\label{subsec:proj}
Another, quite different, idea is to consider only one commodity.
By projecting bid $i$ on commodity $j$, to 
\mbox{$i' =$} 
\mbox{$\langle 0 , \ldots$}, 
\mbox{$q(i,j) , \ldots$},
\mbox{$0 , p(i) \rangle$},
we transform our original MUCA into a knapsack problem.
The optimal solution to the knapsack problem is obviously an upper bound
for the optimal solution of the original MUCA.
The knapsack being polynomially-approximable at any precision, one may
easily obtain an upper bound in this way, in fact $n$ such upper bounds,
one for each commodity.
\subsection{Average price consideration}
\label{subsec:fast}
The bounds described above are not difficult to compute and quite
different in spirit (and probably results).
Our next bound is related to the Linear Programming one of
Section~\ref{subsec:LP}: it always provide an upper bound that is larger than
or equal to the one provided by LP, but it is extremely easy
to compute.
Consider the average price per unit for bid $i$:
\mbox{$a(i) = p(i) \: / \: \sum_{j=1}^{n} q(i,j)$}.
Suppose $a$ is the largest of the $a(i)$'s.
Then clearly the value of the optimal solution to the MUCA is bounded
from above by \mbox{$a \: \sum_{j=1}^{n} k_{j}$}.
The value of the optimal solution to the relaxed LP problem is also
bounded by this quantity.
\section{Choosing the most promising bid}
\label{sec:ordering}
Given positive integers \mbox{$k_{1} , \ldots , k_{n}$},
how should one compare the attractiveness, at first sight, of two bids
\mbox{$a =$} \mbox{$\langle q_{1} , q_{2} , \ldots$}, 
\mbox{$q_{n} , p \rangle$} and
\mbox{$a' =$} \mbox{$\langle q'_{1} , q'_{2} , \ldots$}, 
\mbox{$q'_{n} , p' \rangle$}?
We may as well say we are looking for a criterion $r$, i.e. a real-valued 
function of the bid $a$, possibly depending on the
parameters $k_{j}$'s such that \mbox{$r(a) \geq r(a')$} iff $a$ is more
promising than $a'$.

Some obvious considerations may be made immediately.
The corresponding heuristics have been incorporated in~\cite{Camus2:unpub}.
If the quantities $q_{j}$ and $q'_{j}$ are equal, but $p$ is larger than
$p'$, one should obviously prefer $a$ to $a'$ and in fact, since the two
bids conflict, one can remove $a'$ from consideration.
Similarly, if the prices $p$ and $p'$ are equal and \mbox{$q_{j} \leq q'_{j}$}
for every $j$, \mbox{$1 \leq j \leq n$}, then one should prefer $a$ to $a'$
and remove $a'$ from consideration.
In other terms, the function $r$ should be monotonic in $p$ and anti-monotonic
in the $q_{j}$'s.
Notice that any such function will have the effect of removing from 
consideration all the dominated bids as above.

Many functions come to mind.
The simplest one is probably:
\begin{equation}
\label{eq:p}
r(a) = p \: ,
\end{equation}
i.e., order the bids by the price they propose.
The most natural function that comes to mind is probably:
\begin{equation}
\label{eq:linear}
r(a) = \frac{p}{\sum_{j=1}^{n} q_{j}} \: ,
\end{equation}
that orders the bids by average price per unit.
But one may prefer to normalize the elements of the sum by the number of
units of each commodity that are available and use:
\begin{equation}
\label{eq:linearnorm}
r(a) = \frac{p}{\sum_{j=1}^{n} q_{j} / k_{j}} \: .
\end{equation}
If one has a geometrical bind one may prefer the Euclidean:
\begin{equation}
\label{eq:euclid}
r(a) = \frac{p}{\sqrt{\sum_{j=1}^{n} q_{j}^{2}}} \: ,
\end{equation}
or the normalized:
\begin{equation}
\label{eq:euclidnorm}
r(a) = \frac{p}{\sqrt{\sum_{j=1}^{n} (q_{j} / k_{j})^{2}}} \: .
\end{equation}
A previous result on combinatorial auctions~\cite{LCS:EC99} suggests
the consideration of:
\begin{equation}
\label{eq:unnorm0}
r(a) = \frac{p}{\sqrt{\sum_{j=1}^{n} q_{j}}} \: ,
\end{equation}
or of:
\begin{equation}
\label{eq:norm}
r(a) = \frac{p}{\sqrt{\sum_{j=1}^{n} q_{j} / k_{j}}} \: .
\end{equation}
Many other criteria could be considered, of the form:
\[
r(a) = \frac{p}{(\sum_{j=1}^{n} q_{j}^{l})^{m}} \: 
\]
or
\[
r(a) = \frac{p}{(\sum_{j=1}^{n} (q_{j}/k_{j})^{l})^{m}} \: 
\]
It is extremely difficult to guess what is the best criterion to use
in a branch-and-bound algorithm. Experimental results would be interesting
and we are in the process of obtaining such results but the lack of real-life
data casts a doubt on the applicability of the conclusions that can
be drawn from synthesized data.

Based on each of those criteria, one can devise a (different) 
greedy algorithm to find a solution to MUCAs: pick the most promising
bid in the partial solution, then find the most promising remaining bid 
that does not conflict with the partial solution, and so on until no bid 
can be added. We are looking for the criterion that gives the {\em best}
greedy algorithm. In what sense? 
We choose to compare greedy algorithms
by the approximation ratio they provide in the worst-case.
A greedy algorithm that provides, on any MUCA, a solution
that is at least one-tenth of the optimal solution will be preferred to
any algorithm that provides only one-hundredth of the optimal
solution on some MUCA. Notice that we could assume a probability
distribution on the inputs and compare the expected values of the solutions
found (or their ratio to the optimal one), but this is not what we
propose to do. Our choice is consistent with that of the theoretical 
CS community.
The next section is devoted to theoretical considerations
leading to the characterization of the best criterion.
\section{Theoretical considerations:\newline 
the weighted multi-set packing (WMSP) problem}
\label{sec:theory}
The problem of finding an optimal allocation in a multi-unit combinatorial
auction is best described as a generalization of the weighted set packing
problem~\cite{Halldors:COCOON99,Vemug:Hand}: the weighted multi-set
packing problem. It seems the same problem has been studied 
in~\cite{KanStougieVer:mknap}.

We consider $n$ commodities and assume $k_{j}$ units of each of $n$ 
commodities:
\mbox{$j = 1 , \ldots , n$} are available. 
The $k_{j}$'s are positive natural numbers.
We shall denote by $k$ the quantity \mbox{$\sum_{j=1}^{n} k_{j}$}.
On the whole $k$ items are to (may) be allocated.
A bid is written: \mbox{$\langle q_{1} , q_{2} , \ldots , q_{n} , p \rangle$},
where the $q$'s are natural numbers with \mbox{$q_{j} \leq k_{j}$}
and $p$ is a natural number (or a rational nonnegative number).
A bid is an offer to acquire $q_{j}$ units of each commodity $j$ for
a total sum of $p$.
Given a set of bids we want to find a subset that maximizes the sum of the
$p$'s such that for every $j$, \mbox{$1 \leq j \leq n$}, the sum of the
$q_{j}$'s is less than or equal to $k_{j}$.
A number of special cases are worth noticing:
\begin{itemize}
\item if all $k_{j}$'s are equal to one, the problem is 
the weighted set packing (WSP) problem,
\item if there is only one commodity ($n=1$), the problem is the knapsack
problem.
\end{itemize}

Our first result, a trivial generalization from~\cite{Sandholm99,LCS:EC99}, 
shows that the
WMSP problem is not only NP-hard but also hard to approximate.
\begin{theorem}
\label{the:inapprox}
Unless NP=ZPP\footnote{A language $L$ belongs to ZPP if and
only if there is some constant $c$ such that there is a probabilistic
Turing machine $M$ that on input $x$ runs in expected time
\mbox{$O({\mid x \mid}^{c})$} and outputs $1$ if and only if
\mbox{$x \in L$}}, the WMSP problem cannot be approximated 
within $n^{1/2 - \epsilon}$
in polynomial time, for any \mbox{$\epsilon > 0$}. 
\end{theorem}
\begin{proof}
A graph with $v$ vertices and $e$ edges defines a WMSP problem 
in the following way.
Consider $e$ commodities (\mbox{$n = e$}) and assume $k_{j}$ units
of commodity $j$ are for sale. 
The $k_{j}$'s are arbitrary.
There are $v$ bids.
Bid $i$ offers a price of $1$ for $k_{j}$ units of each of the commodities
(i.e. edges) $j$ that vertex $i$ is adjacent to.
Any feasible allocation defines a set of independent vertices and its value is
the size of the independent set. The solution of the WMSP problem is therefore
equivalent to finding a maximal independent set. 
An $f(n)$-approximation of the WMSP problem provides 
an $f(e)$-approximation of the
Maximal Independent Set problem. 
By~\cite{Hastad:CliqueActa}, no efficient $k^{1 - \epsilon}$-approximation
exists for the maximal independent set problem, 
therefore no $e^{1/2 - \epsilon}$-approximation, unless NP=ZPP.
\end{proof}
In~\cite{LCS:EC99} it is shown that the WSP problem admits a polynomial-time 
$n^{1/2}$-approximation and that there is a greedy algorithm that
achieves this (optimal) approximation ratio. 
Notice that finding a polynomial-time 
$k^{-1}$-approximation of the Maximal Independent Set is trivial, but finding
such an $n^{1/2}$-approximation for WSP is not.
In the more general case of WMSP problems, we shall characterize the
approximation quality of greedy algorithms as 
\mbox{$\sqrt{k} = \sqrt{\sum_{j=1}^{n} k_{j}}$} in Theorems~\ref{the:lowerb} 
and~\ref{the:upperb},
but we do not know how to match the lower bound of Theorem~\ref{the:inapprox}.
For reasons explained above, we are interested in solving (approximately)
WMSP problems by a greedy algorithm. For such algorithms, we may prove a 
result that is stronger than Theorem~\ref{the:inapprox}.
\begin{theorem}
\label{the:lowerb}
No (polynomial-time) greedy algorithm can guarantee an approximation of better
than \mbox{$\sqrt{k}$}.
\end{theorem}
\begin{proof}
Define a {\em unit} bid as a bid offering a sum of $1$ for a single unit of a
single commodity: \mbox{$\langle 0 , 0 , \ldots , 1 , \ldots , 0 , 1 \rangle$}.
There are $n$ different unit bids, but the same unit bid may appear a number
of times in the list of bids. 
Given a greedy algorithm, let $u$ be the unit bid that is ranked first 
among all unit bids by the algorithm.

Two problems, i.e., sets of bids, will be described.
A greedy algorithm will perform badly, i.e. propose a solution
that is only a $\sqrt{k}$-approximation, 
on one of those two problems.
In the first problem, there are only two bids:
\begin{enumerate}
\item 
a bid for all the units available:
\mbox{$A =$} \mbox{$\langle k_{1} , k_{2} , \ldots$}, 
\mbox{$k_{n} , \sqrt{k} \rangle$}, and
\item
a bid $u$.
\end{enumerate}
In the second situation, there are \mbox{$k  + 1$} bids:
\begin{enumerate}
\item 
a bid $A$ for all the units available as above, and
\item
for every $j$, \mbox{$j = 1 , \ldots , n$}, $k_{j}$ unit bids for
commodity $j$:
\mbox{$\langle 0 , 0 , \ldots , 1 , \ldots , 0 , 1 \rangle$}.
\end{enumerate}
If bid $A$ is ranked before $u$, then, in the second situation,
the greedy method obtains \mbox{$\sqrt{k}$} instead of the optimal
\mbox{$k$}, a \mbox{$\sqrt{k}$}-approximation. 
If bid $u$ is ranked before bid $A$, then,
in the first situation the greedy method obtains $1$ instead of the 
optimal \mbox{$\sqrt{k}$}, again a \mbox{$\sqrt{k}$}-approximation.
For any greedy method, one of the two situations will give a solution
that is a factor of \mbox{$\sqrt{k}$} less than optimal. 
\end{proof}

The following positive result provides an upper-bound 
that matches the
lower bound of Theorem~\ref{the:lowerb}.
\begin{theorem}
\label{the:upperb}
A greedy algorithm (to be described) provides a polynomial-time
$\sqrt{k}$-approximation for the WMSP problem.
\end{theorem}
\begin{proof}
We shall use the index $i$ to range over the bids.
The bid $i$ is:
\mbox{$i = \langle q(i,1) , q(i,2) , \ldots , q(i,n) , p(i) \rangle$}.
We define \mbox{$w(i) = \sqrt{\sum_{j=1}^{n} q(i,j)}$}.
The algorithm we propose is greedy allocation based on the following
criterion $r$ to rank the bids, in descending order:
\begin{equation}
\label{eq:unnorm}
r(i) = \frac{p(i)}{w(i)}
\end{equation}
Let $OP$ be the optimal solution, i.e., the set of bids contained
in the optimal solution.
The value of the optimal solution is \mbox{$\alpha = \sum_{i \in OP} p(i)$}.
Let $GR$ be the solution obtained by the greedy allocation and $\beta$
its value: \mbox{$\beta = \sum_{i \in GR} p(i)$}.
We want to show that:
\begin{equation}
\label{eq:goal}
\alpha \leq \beta \: \sqrt{k}.
\end{equation}
Notice, first, that we may, without loss of generality, assume that
the sets $OP$ and $GR$ have no bid in common.
Indeed, if they have, one considers the problem in which the common bids and
all the units they request have been removed. The greedy and optimal solutions
of the new problem are similar to the old ones and the inequality for the new smaller
problem implies the same for the original problem.

Let us consider $\beta$.
By elementary algebraic considerations:
\[
\beta = \sum_{i \in GR} p(i) \geq \sqrt{\sum_{i \in GR} {p(i)}^{2}} =
\sqrt{\sum_{i \in GR} {r(i)}^{2} \: {w(i)}^{2}} =
\]
\[
\sqrt{\sum_{i \in GR} {r(i)}^{2} \: \sum_{j=1}^{n} q(i,j)} = 
\sqrt{\sum_{j=1}^{n} \sum_{i \in GR} {r(i)}^{2} \:  q(i,j)}.
\]
Consider $\alpha$.
By the Cauchy-Schwarz inequality:
\[
\alpha = \sum_{i \in OP} r(i) \: w(i) \leq \sqrt{\sum_{i \in OP} {r(i)}^{2}}
\: \sqrt{\sum_{i \in OP} {w(i)}^{2}}.
\]
But:
\[
\sum_{i \in OP} {w(i)}^{2} = \sum_{i \in OP} \sum_{j=1}^{n} q(i,j) =
\sum_{j=1}^{n} \sum_{i \in OP} q(i,j).
\]
The expression \mbox{$\sum_{i \in OP} q(i,j)$} represents the total number
of units of commodity $j$ allocated in the optimal allocation $OP$
and is therefore bounded from above by $k_{j}$, the number of units
available.
We conclude that:
\[ 
\alpha \leq \sqrt{\sum_{i \in OP} {r(i)}^{2}} \: \sqrt{k}.
\]

To prove~(\ref{eq:goal}), it will be enough, then, to prove that:
\[
\sum_{i \in OP} {r(i)}^{2} \leq 
\sum_{j=1}^{n} \sum_{i \in GR} {r(i)}^{2} \:  q(i,j).
\]

Consider the optimal solution $OP$. By assumption, the bids of $OP$ did not
enter the greedy solution $GR$. This means that, at the time such a bid $i$
is considered during the execution of the greedy algorithm, it cannot
be entered in the partial allocation already built.
This implies that there is a commodity $j$, not enough units of which
are still unallocated to satisfy the quantity $q(i,j)$ requested by
bid $i$. In other terms the sum of $q(i,j)$ and the $q(*,j)$ of all the bids
of the greedy solution already considered was larger than $k_{j}$:
\[
q(i,j) + \sum_{l \in GR , r(l) \geq r(i)} q(l,j) \: > \: k_{j}.
\]
In particular it follows that \mbox{$q(i,j) > 0$} and therefore 
\mbox{$q(i,j) \geq 1$}.
We attach to every bid $i$ of $OP$ such a commodity $c(i)$.
If more than one such suitable commodity exists, we choose one arbitrarily.
Let $OP_{j}$ be the subset of $OP$ that contains those bids $i$ for which
\mbox{$c(i) = j$}. The $OP_{j}$ provide a partition of $OP$ and:
\[
\sum_{i \in OP} {r(i)}^{2} = \sum_{j=1}^{n} \sum_{i \in OP_{j}} {r(i)}^{2}.
\]
Let us denote by $m_{j}$ the size of the set $OP_{j}$.
We shall conclude the proof by showing that, for every $j$, 
\mbox{$1 \leq j \leq n$}:
\[
\sum_{i \in OP_{j}} {r(i)}^{2} \leq \sum_{i \in GR} {r(i)}^{2} \:  q(i,j).
\]
Let $l_{j}$ be a bid of $OP_{j}$ whose $r$ is maximal (among bids of $OP_{j}$)
and let \mbox{$r_{j} = r(l_{j})$}.
We know that:
\[
\sum_{i \in OP_{j}} {r(i)}^{2} \leq {r_{j}}^{2} \: m_{j}.
\]
We want to bound $m_{j}$ from above.
We remarked above that:
\[
q(l_{j},j) \: + \: \sum_{l \in GR , r(l) \geq r_{j}} q(l,j) > k_{j}.
\]
and that \mbox{$q(l,j) \geq 1$} for every \mbox{$l \in OPT_{j}$}.
Therefore:
\[
k_{j} \: \geq \: \sum_{l \in OP_{j}} q(l,j) \: \geq \: q(l_{j},j) + m_{j} - 1 
\: > 
\]
\[
\: k_{j} - \sum_{l \in GR , r(l) \geq r_{j}} q(l,j) \: + m_{j} - 1
\]
and therefore \mbox{$m_{j} \leq \sum_{l \in GR , r(l) \geq r_{j}} q(l,j)$}.
Clearly, then:
\[
{r_{j}}^{2} \: m_{j} \: \leq \:
{r_{j}}^{2} \: \sum_{l \in GR , r(l) \geq r_{j}} q(l,j) \: \leq 
\]
\[
\sum_{l \in GR , r(l) \geq r_{j}} {r(l)}^{2} \: q(l,j) \: \leq \:
\sum_{l \in GR} {r(l)}^{2} \: q(l,j)
\]
\end{proof}
Theorems~\ref{the:lowerb} and~\ref{the:upperb} show
that the criterion of Equation~\ref{eq:unnorm0} 
is optimal, in a worst-case sense.
\begin{corollary}
\label{co:optimal}
The criterion proposed in Equation~\ref{eq:unnorm0} for choosing the
most promising bid guarantees, in the worst-case, the best possible 
approximation ratio.
\end{corollary}
None of the other criteria examined in Section~\ref{sec:ordering}
achieves the same approximation ratio.
Let us show this explicitly for the normalized criterion of 
Equation~\ref{eq:norm}, and characterize exactly the approximation
ratio achieved by this criterion.

Along the lines of the proof of Theorem~\ref{the:upperb}, one can show 
that using the normalized ranking criterion:
\[
r(i) = p(i) / \sqrt{\sum_{j=1}^{n} q(i,j) \: / \: k_{j}}
\]
achieves a $\sqrt{M n}$-approximation where $M$ is the maximum of the
$k_{j}$'s. 
The following example shows that this upper bound is exact.
Therefore the normalized criterion is not optimal.

Consider two commodities, and assume there are $k$ units of the first one
(\mbox{$k_{1} = k$}) and one unit of the second one (\mbox{$k_{2} = 1$}).
There are two bids. The first one is \mbox{$\langle k , 1 , \sqrt{2} \rangle$}
and the second one is \mbox{$\langle 1 , 0 , 1 / \sqrt{k} \rangle$}.
Both bids have the same normalized ranking criterion ($1$). 
Assume the greedy method places the second bid first.
It obtains $1 / \sqrt{k}$ instead of the optimal $\sqrt{2}$:
an $\sqrt{2 \: k}$-approximation, not as good as the $\sqrt{k + 1}$-approximation
provided by the unnormalized criterion.
In Section~\ref{sec:BB}, we indicated that we were willing to consider
dynamic, and not only static, ordering criteria for the bids.
Our conclusion is that the optimal criterion is a static one, 
quite a surprising conclusion.
\section{Practical considerations for \\choosing the most promising bid}
\label{sec:pract}
As noted in Section~\ref{sec:bounding} the choice of a method for
bounding the possible value of an auction from above is quite
unproblematic in practice: one may use a wealth of methods 
and take the smallest upper-bound obtained.
In practice, one will, on the basis of the record of past runs, 
easily find out which of the methods are useless and stop using them.
The choice of the most promising bid studied in Section~\ref{sec:ordering}
and~\ref{sec:theory} is much more of a problem.
The criterion described in~(\ref{eq:unnorm0}) is shown to be 
theoretically optimal in Corollary~\ref{co:optimal},
but is it the best in practice?
It is the criterion that guarantees the best approximation ratio in
the worst case, does it provide the best solution in practice?
The truth of the matter is that only experience with real MUCAs can tell
and, at this moment, no such data exists.
We can only point out at two considerations.
First, the examples of Section~\ref{sec:theory} present
quite tellingly why the criterion defined in~(\ref{eq:unnorm0})
strikes a balance between a criterion favoring large bids and a global
view such as the
one defined in~(\ref{eq:p}) or one that favors small bids and fine tuning
such as defined in~(\ref{eq:linear}) or in~(\ref{eq:euclid}).
The proof of Theorem~\ref{the:upperb} also shows why the unnormalized
criterion of~(\ref{eq:unnorm0}) should be preferred to the normalized criterion
of~(\ref{eq:norm}).
Secondly, we have confidence in the applicability of theoretical results:
techniques that can be proved to be optimal in theory tend to work well.

In~\cite{Camus2:unpub}, the authors use a complex method for choosing
the most promising bid: their choice depends on the other bids and also
on the results of the bounding-from-above procedure.
Since the ordering of the commodities does not depend on the
price offered by the bids, but may determine the solution
that will be considered first, it seems that this first candidate
solution may be arbitrarily worse than the optimal solution.
The basic criterion used inside bins is similar to that of (\ref{eq:linear}) 
and therefore not optimal in theory.
\section{Experimental results}
Experimental results are crucial in the assessment of the ideas developed
above. Three basic questions should be answered.
\begin{enumerate}
\item Does the criterion for choosing the most promising bid that
has been shown to be optimal in Corollary~\ref{co:optimal} perform
well in practice, i.e., does a search algorithm that uses this criterion
find rapidly the optimal solution?
\item Do the different methods for bounding from above the value of the
optimal solution lead to the pruning of many large subtrees?
Which of those methods are most useful?
\item Are those bounding methods fast enough to be usable in practice
or does their use imply that the algorithm spends an unbearable amount
of time in computing upper bounds?
A trade-off between the amount of effort spent in pruning and in examining
new nodes must be struck.
\end{enumerate}
The best would obviously be to examine those questions on data
obtained from real-life combinatorial auctions.
Such data is very difficult to find and we could not put our hands
on such data.
The next best thing that can be done is to use auctions artificially generated.
In~\cite{Camus2:unpub}, the authors define a probability distribution over
combinatorial auctions and test the average behavior of their algorithm on 
this distribution. We chose to use the same distribution, with the same
values for the parameters.

Let us address the second question first. The answer is emphatically 
positive. The methods proposed above provide upper bounds that allow 
for an extremely thorough pruning of the search tree.
Linear Programming provides a very tight upper bound that leads to
a very short search path over the tree: only a very small fraction
of the nodes have to be expanded. For example, considering $40$ bids,
we found that only $1.6 \times {10}^{-9}$ of the nodes were visited.

The projections bounds are not as good, and provide, for the distribution
we used, bounds that are not better than the bound provided by
average price considerations.
This is probably due to the specific distribution chosen, 
in which it is rarely the case that a single good is {\em dominant}, 
such as, e.g., in the case all bids request all the units of a
certain specific good.
One of the characteristics of the distribution suggested 
in~\cite{Camus2:unpub} is that bids tend to request only a small number of 
units per good. This kind of distribution leads 
to an auction with no dominant good for which the projection bounds
are quite loose. R. Holte~\cite{Holte:private} found that, on another
distribution, the projection bound is very good.
Nevertheless, projection bounds allow us to visit only 
$1.6 \times {10}^{-8}$ of the nodes.

The average price upper bound is not as good as the Linear Programming
one, but is, on average, as good as the projection bound.
It allows for a very thorough pruning of the search tree. 
Considering $40$ bids,
we found that only $1.6 \times {10}^{-8}$ of the nodes had to be visited.
This number is the same as the one for the projection bound and ten times
larger than the one for Linear Programming.

Let us now address the third question. We found Linear Programming
to be extremely costly (in time) to use, so costly as to render its use
infeasible for large combinatorial auctions. 
Since we have only just begun experimenting, we hope that we shall learn
in the future how to use Linear Programming effectively.
The average price upper bound is, on the contrary, computed very fast.
The projection bounds are not computed as fast.
Since we explained above that the projection bounds were, on our
distribution, not better, 
{\em the results to be presented below have all been obtained by using
the average price upper bound exclusively.}
\begin{figure}
\centering
\epsfig{file=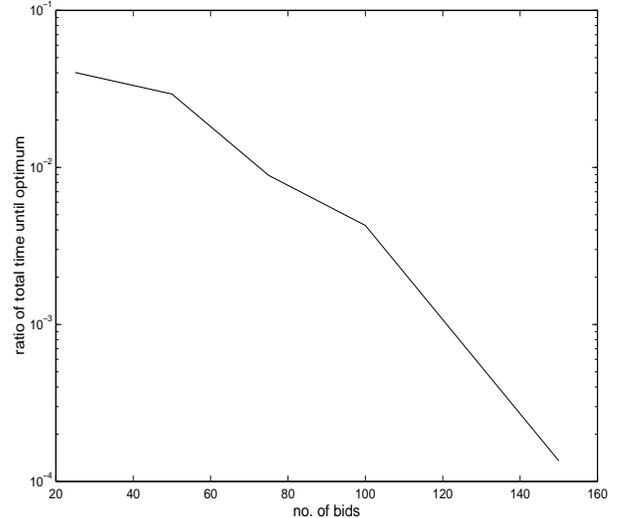,height=7cm, width=8cm}
\caption{This graph shows the part of the execution time 
spent until the optimal solution is found as a function of 
the number of bids for auctions of fourteen goods.
Auctions are distributed as proposed by Leyton-Brown, Shoham and Tennenholtz.
Times are averaged over ten trials.}
\end{figure}

Let us discuss the first question now.
The best test of the quality of our ordering heuristic is probably 
to examine how fast the optimal solution is obtained.
We found that, using the ordering criterion described in 
Equation~\ref{eq:unnorm0}, the optimal solution is found extremely 
rapidly and the algorithm spends an overwhelming part of its time in
showing that this solution is indeed optimal.
In Figure~1, we plot the time spent until the optimal solution
is found divided by the total time spent in the search
for auctions of different sizes. All auctions include $14$ goods,
the number of bids is found on the $x$ axis.
For small auctions this number is of the order of $4\%$;
for larger auctions this number decreases about linearly on a logarithmic
scale and for auctions of $150$ bids, the optimal solution is found
after only $0.014\%$ of the total execution time.
It seems that this percentage decreases at least exponentially.
Those results seem to improve significantly on those of~\cite{Camus2:unpub}.
The {\em Percentage Optimality} graph there seems to indicate than more than
$10\%$ of the time is spent before the optimal solution is found.
Those results indicate that the criterion of Equation~\ref{eq:unnorm0}
that we proved theoretically optimal is extremely good in practice.
\begin{figure}
\centering
\epsfig{file=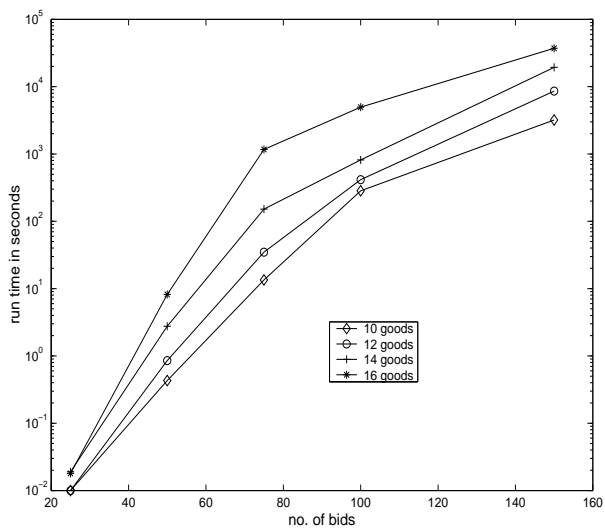, height=7cm, width=8cm}
\caption{This graph shows the execution time for auctions containing
different numbers of bids.
Auctions are distributed as proposed by Leyton-Brown, Shoham and Tennenholtz.
Times are averaged over ten trials.
Each of the four curves corresponds to a different number of goods.}
\end{figure}

Figure~2 describes the execution times of our branch and bound 
algorithm using the average price upper bound and the ordering heuristic
of Equation~\ref{eq:unnorm0}:
on the $x$-axis the number of bids, on the $y$-axis the execution time.
Four different curves are plotted, corresponding to a different number 
of goods.
Our experimental data was collected on a Pentium III-450 running
Linux, using 640K of memory.
Figure~2 is comparable to the {\em Number of Bids vs. Time}
figure in~\cite{Camus2:unpub}.
Our running times are larger, by orders of magnitude, than theirs.
For this reason, we were not even capable of solving the smallest auctions
they considered.
Figure~2 shows very clearly a sub-linear (on a logarithmic scale)
growth in execution time, indicating that the growth is sub-exponential.
This feature is also found, but less clearly, in~\cite{Camus2:unpub}'s graph.
We intend to look into the reasons for the huge gap in running times
between our algorithm and CAMUS. The huge discrepancy between the amount
of memory they use (25 MB) and ours (640K) is certainly part of 
the explanation.
\begin{figure}
\centering
\epsfig{file=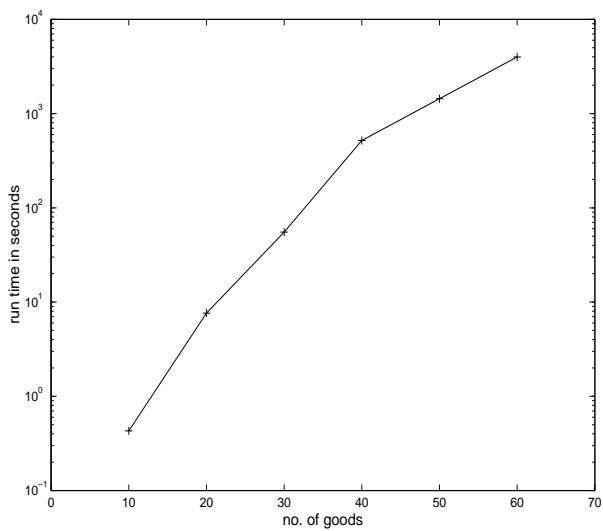, height=7cm, width=8cm}
\caption{This graph shows the execution time for 50 bids auctions with
different numbers of goods.
Auctions are distributed as proposed by Leyton-Brown, Shoham and Tennenholtz.
Times are averaged over ten trials.}
\end{figure}

Figure~3 examines the dependence of the running time on
the number of goods. The number of bids is fixed at $50$, and
the number of goods is described on the $x$ axis.
CAMUS~\cite{Camus2:unpub} exhibits an exponential sensitivity to the
number of goods (as opposed to the number of bids).
Figure~3 clearly shows a sub-exponential growth.
\section{Conclusions and further work}
\label{sec:conc}
We proposed a simple branch-and-bound framework to solve MUCAs.
We succeeded in characterizing the theoretically optimal method for
sorting bids.
Two main directions for research are left open:
\begin{itemize}
\item characterize the power of different methods for bounding from above
the value of MUCAs
\item run more extensive experiments to compare the performance of 
different heuristics
both for bounding and for ordering and to compare our proposal to others
such as in~\cite{Camus2:unpub}.
\item consider more efficient ways of using Linear Programming,
using the results of previous computations 
to speed up the search for a solution.
\end{itemize}
\section{Acknowledgements}
Conversations with Michel Bercovier and Michael Ben-Or 
and remarks from Amir Ronen
are gratefully acknowledged.
The first author is partially supported by Grant 15561-1-99 from the Israel
Ministry of Science, Culture and Sport and by 
the Jean and Helene Alfassa fund for 
research in Artificial Intelligence.
\bibliographystyle{abbrv}

\end{document}